\documentclass[10pt,letterpaper]{article}
\usepackage{opex3}
\begin{document}

\title{Carrier-envelope phase measurement from half-cycle high harmonics}
\author{Pengfei Lan, Peixiang Lu$^\dag$, Fang Li, Yuhua Li, Zhenyu Yang}
\address{Wuhan National Laboratory for Optoelectronics, Huazhong University of
Science and Technology, Wuhan 430074, P. R. China}
\email{lupeixiang@mail.hust.edu.cn}

\begin{abstract}
We present a method to distinguish the high harmonics generated in
individual half-cycle of the driving laser pulse by mixing a weak
ultraviolet pulse, enabling one to observe the cutoff of each
half-cycle harmonic. We show that the detail information of the
driving laser pulse, including the laser intensity, pulse duration
and carrier-envelope phase, can be {\it in situ} retrieved from
the harmonic spectrogram. In addition, our results show that this
method also distinguishes the half-cycle high harmonics for a
pulse longer than 10 fs, suggesting a potential to extend the CEP
measurement to the multi-cycle regime.
\end{abstract}

Nowadays, rapid advances in laser technology have made it possible
to shape and control intense laser pulses consisting only of a few
optical cycles (typically, 5 fs for the Ti:sapphira laser),
opening up new frontiers for ultrafast physics \cite{T.Brabec}. To
fully characterize such few-cycle pulses, the relative phase of
the carrier wave with respect to the pulse envelope, i.e., the
carrier-envelope phase (CEP) is required besides the pulse
intensity, envelope and wavelength. In the few-cycle regime, the
temporal evolution of the electric field depends sensitively on
the CEP. Consequently, many strong-field interactions which are
relevant to the electric field of the pulse rather than the
intensity profile, become sensitive on the CEP
\cite{A.Baltuska,D.B.Milosevic}. Then the phase-stabilized
few-cycle laser pulses provide us a precise tool to coherent
control the electronic processes \cite{A.Baltuska}, and the
ability to measure and stabilize the CEP therefore becomes a
crucial point for all applications.

Many methods have been proposed for CEP measurement based on the
photonionization
\cite{D.B.Milosevic,G.Paulus,A.D.Bandrauk,A.Apolonski,C.Lemell},
ultraviolet (uv) and terahertz emission
\cite{C.A.Haworth,M.Kress}. It was shown that the few-cycle laser
pulse leads to a strong left-right asymmetry of photoelectrons
\cite{D.B.Milosevic}, which has been observed and utilized as a
meter of the CEP \cite{G.Paulus}. The photonelectron emission from
metal surfaces also shows a phase-sensitive property
\cite{A.Apolonski}, suggesting a potential to measure the CEP
\cite{C.Lemell}. But for photonelectron signals, thousands of
consecutive pulses measurement is usually required to achieve the
required degree of accuracy. In recent work, Haworth {\it et al.}
proposed a method of spatio-filtering, which separated the high
harmonic generation (HHG) in each half-cycle of the laser pulse in
the spatio domain and the CEP was retrieved form the harmonic
spectrogram. Moreover, the high harmonic signal is easily detect
and single short measurement becomes possible. In addition, the
spatio-filtering indicates a potential to CEP measurement for a
multi-cycle pulse ($\ge$10 fs), i.e., better by half
\cite{S.T.Cundiff}.

\begin{figure}[htbp]
\centering\includegraphics[width=11cm]{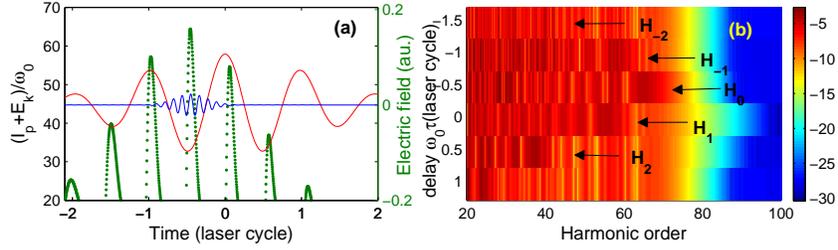}
\caption{\label{fig1} (a) Electric fields of the few-cycle laser
pulse (red line) and the uv pulse (blue line). The dotted line
shows the dependence of the electron energy ($I_p+E_k$) on the
ionization time in the few-cycle laser pulse. (b) High harmonic
spectrogram as a function of delay between the laser and uv pulses
(see the movie for various CEPs). The colorbar shows the harmonic
intensity in logarithmic scale. The laser intensity is
$4\times10^{14}$ $\textrm{Wcm}^{-2}$, pulse duration is $5$ fs
full width at half maximum and the CEP is 0. The uv pulse
intensity is $1\times10^{13}$ $\textrm{Wcm}^{-2}$ and the duration
is $1$ fs.}
\end{figure}

In this work, we propose an alternative method to separate the
half-cycle harmonic (HCH) in the time domain, from which we
retrieve the information of driving laser pulse, including the
laser intensity, pulse duration and CEP. Our method is based on
the HHG in a two-color field, which is illustrated in Fig.
\ref{fig1}(a). The red line shows the 5-fs laser field $E(t)$. As
the three-step model \cite{P.Corkum}, the electron is first set
free via photonionization at the laser peak and then is
accelerated in the following half-cycle of the laser field,
accumulating a kinetic energy of
$E_k=(\int_{t_i}^{t_r}E(t)\textrm{d}t)^2/2$ with a maximum of
about $3.17U_p=3.17E_{0}^2/(4\omega_0^2)$, where $t_i$, $t_r$ are
the ionization and recombination times, $U_p$ is the ponderomotive
energy and $E_{0}$ is the laser amplitude. Finally, the electron
recombines with the parent ion near the zero transition of the
laser field by releasing its energy to high harmonics with the
cutoff energy of $I_p+3.17U_p$, where $I_p$ is the atomic
ionization energy. Such a process occurs periodically in each
half-cycle of the driving field as shown by the dotted line in
Fig. \ref{fig1}(a). However, how to observe the HCH still is a
challenge in experiments. As shown in Fig. \ref{fig1}(a),
ionization is the starter of HHG and each HCH corresponds to a
specific ionization time near the laser peak. Therefore, we mix a
uv pulse to the laser pulse (see the blue line). Due to the high
photon energy and ultrashort duration, the uv pulse significantly
enhances the photonionization \cite{K.Ishikawa} and also confines
the ionization time in an interval of several hundred attoseconds
\cite{K.J.Schafer,P.Lan}, which has been demonstrated to control
and enhance the HHG. Here, we focus on how to distinguish the
individual HCH and to extract the driving pulse information from
the harmonic spectrogram. In the two-color field as shown in Fig.
\ref{fig1}(a), the uv pulse firstly launches a free electron,
initialing the HHG. The electron is then accelerated in the
following half-cycle of the driving laser field, and finally
recombines with the parent ion by releasing high harmonics. Since
the ionization is enhanced and confined in a short interval by the
uv pulse, high harmonics generated in a half-cycle can be
selected. Scanning the uv pulse, each HCH will be separated in the
time domain, which also inherits the information of the driving
field. From the harmonic spectrogram, the laser intensity, pulse
duration and CEP are therefore can be {\it in situ} retrieved
during HHG.

For simulation, we solve the time-dependent Sch\"odinger equation
(TDSE) with the split-operator method \cite{P.Lan,M.Feit}. We
choose a Ti:sapphire 5-fs laser pulse with a wavelength of 800 nm
and an intensity of $4\times10^{14}$ $\textrm{Wcm}^{-2}$, the
electric field reads $E(t)=E_0f(t)\cos(\omega_0t+\phi_0)$ where
$E_0$ is the laser amplitude, $\omega_0$ and $\phi_0$ are the
laser frequency and CEP, respectively, $f(t)=\exp(-t^2/T^2)$ is
the pulse envelope and the pulse duration is $2\sqrt{ln2/2}T$ full
width at half maximum (FWHM). For the uv pulse,
$E_{uv}(t)=E_{uv0}f(t-\tau)\cos[\omega_{uv}(t-\tau)]$ where $\tau$
is the delay between the laser and uv pulses. The intensity of the
uv pulse is $1\times10^{13}$ $\textrm{Wcm}^{-2}$, wavelength is
100 nm. We intend to select the HCH, and hence set the uv pulse
duration as 1 fs $\sim{T_0}/2$ where $T_0$ is the optical cycle of
the driving field. The harmonic spectrogram is shown in Fig.
\ref{fig1}(b). \begin{figure}[htbp]
\centering\includegraphics[width=11cm]{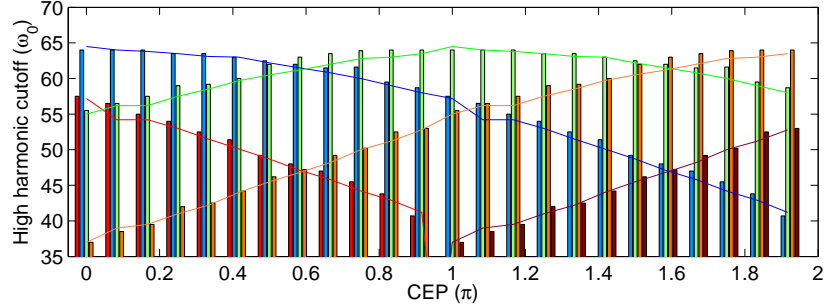}
\caption{\label{fig2} The dependence of half-cycle harmonic
cutoffs on the CEP. Solid lines show the cutoffs obtained with the
classical model. Parameters are the same as Fig. \ref{fig1}.}
\end{figure}One can see that the high harmonics produced in
each half-cycle are clearly separated at different ionization
times and the half-cycle cutoffs (HCOs) are also distinguished.
For instance, at $\omega_0\tau=-0.5T_0$, the harmonic spectrum
(marked by $H_0$) clearly shows a cutoff at $64\omega_0$,
corresponding to the HCH with the ionization time of $-0.5T_0$
shown in Fig. \ref{fig1}(a). Scanning the uv pulse, we observe the
other HCHs, see harmonics $H_{\pm1}$ for the HCH with the
ionization time of $-T_0$ and 0, respectively. In contrast to the
spatio-filtering, our method separates each HCH in the time
domain, which is therefore named as temporal-filtering. Moreover,
our method also

\begin{figure}[htbp]
\centering\includegraphics[width=7cm]{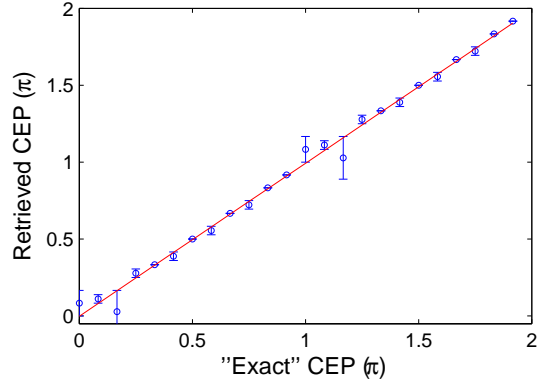}
\caption{\label{fig3} Retrieved CEPs as a function of ``exact''
ones. Parameters are the same as Fig. \ref{fig1}.}
\end{figure}
\noindent clearly separates the HCHs with close cutoffs, e.g.,
$H_{-1}$ and $H_1$, improving the resolution.

For the few-cycle pulse, the laser envelope and hence the
amplitude vary significantly in a half-cycle of the laser pulse,
leading to a different electron energy accumulated in each
half-cycle, which is inherited by the HCH. Conversely, we can
retrieve the detail information of the laser pulse from the
spectrogram. In Fig. \ref{fig2}, we show the half-cycle cutoff
(HCO) as a function of CEPs. For comparison, we also calculated
the HCOs with the classical three-step model \cite{P.Corkum},
modified by taking account the initial energy when the electron is
released. It can be estimated as $\omega_{uv}-I_p-V(t_i)$, where
$V(t_i)$ is the height of the combined Coulomb and driving laser
field barrier at the ionization time \cite{K.J.Schafer}. Note that
the initial energy is not significant for our parameters and is
negligible if the uv photon energy is less than 8 eV (the central
wavelength is about 135 nm). The results are shown by the solid
lines in Fig. \ref{fig2}, which agree quite well with the HCOs
from TDSE. In addition, one can see that the HCO depends
sensitively on the CEP of the driving field, enabling one to
retrieve the CEP from the harmonic spectrogram. For illustration,
we refer the harmonic spectrogram obtained from TDSE to the
``experimental'' results without knowing the CEP, and then present
an algorithm to retrieve the driving field information from the
``experimental'' spectrograms. Firstly, we calculate the HCOs by
the classical model for a 5-fs pulse with a range of intensities
($3.5-4.5\times10^{14}$ $W/cm^2$) and CEPs ($0-2\pi$),
establishing a database of the ``theoretical'' HCOs. Afterward, we
extract the HCOs from the ``experimental'' spectrograms and then
compare it with the ``theoretical'' HCOs. Each comparison returns
a value proportional to how well the two agree, which is defined
as the square of the offset between the ``experimental'' and
``theoretical'' HCOs. Finally, the driving laser intensity and CEP
can be taken from the best match and the error from the standard
deviation of the degree of agreement from this value. The
intensity returned by the algorithm is $4.1\pm0.1\times10^{14}$
$\textrm{Wcm}^{-2}$, which agrees quite well with the ``exact''
value. Figure \ref{fig3} presents the retrieved CEPs, which also
agree well with the ``exact'' ones. The largest deviation is $0.04
\pi$ and the average deviation is only $0.01 \pi$.

\begin{figure}[htbp]
\centering\includegraphics[width=11cm]{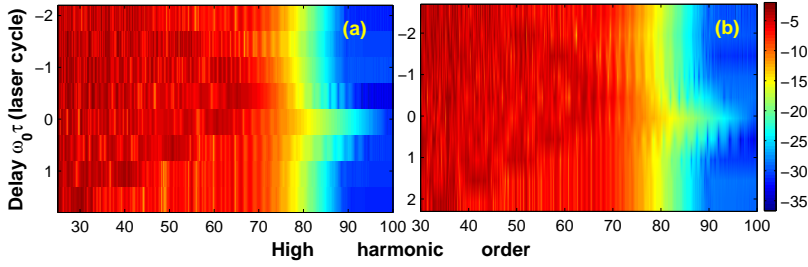}
\caption{\label{fig4} High harmonic spectrogram as a function of
delay between the driving laser and uv pulse in a (a) 8- and
11.5-fs laser pulse. See the movie for a range of CEPs. Other
parameters are the same as Fig. \ref{fig1}.}
\end{figure}

\begin{figure}[htbp]
\centering\includegraphics[width=11cm]{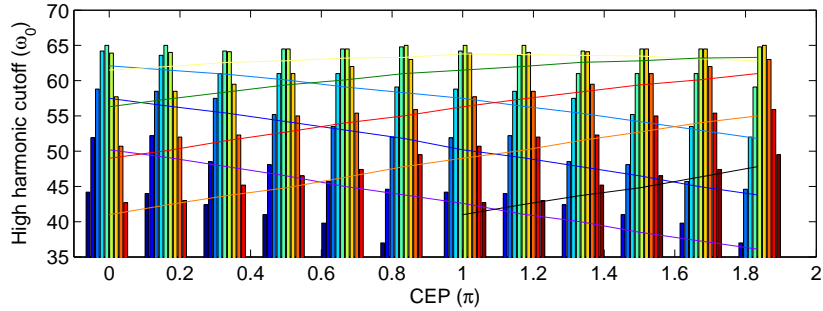}
\caption{\label{fig5} Same as Fig. \ref{fig2}, but for a 11.5-fs
laser pulse.}
\end{figure}

Figure \ref{fig4} shows the harmonic spectrograms as a function of
delay for a (a) 8- and (b) 11.5-fs laser pulses. One can clearly
distinguish each HCH and HCO from these spectrograms. In Fig.
\ref{fig5}, we show the HCOs for a range of CPEs obtained from
TDSE and classical model. Similarly, the detail information of the
driving field can be retrieved from the harmonic spectrogram as
the above discussions. Hence, our method also indicates a
potential to extend the CEP measurement to the multi-cycle regime.

\begin{figure}[htbp]
\centering\includegraphics[width=10cm]{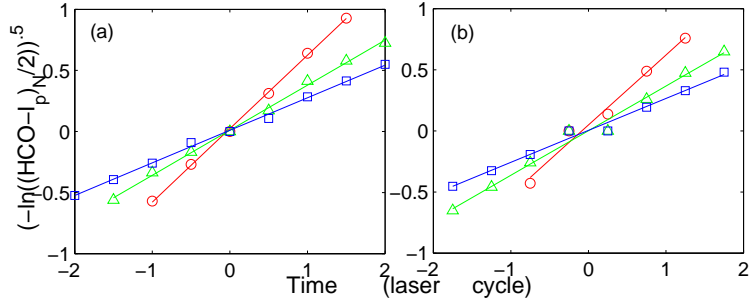}
\caption{\label{fig6} The half-cycle harmonic cutoff
$\sqrt{-ln[(H_c-I_p)_N]/2}\sim{t/T}$ for a two- (circles), three-
(triangle), four-(square) laser pulse with the CEP of (a) 0 and
(b) $\pi/2$. The solid line presents the fitted results. Other
parameters are the same as Fig. \ref{fig1}.}
\end{figure}

Pulse duration is another important parameter for the ultrashort
pulse, which is usually measured by frequency-resolved optical
gating. In principle, the pulse duration can be retrieved from the
comparison algorithm mentioned above, which however enlarges the
calculation. Here we present an alternative method for {\it in
situ} pulse duration measurement during HHG. As the three-step
model, the half-cycle harmonic cutoff
$\textrm{HCO}{\simeq}I_p+3.17(E_{p}/2/\omega_0)^2$, where $E_{p}$
is the laser amplitude of the individual half cycle, i.e., a
series of values on the Gaussian envelope. In other words,
$\textrm{HCO}-I_p{\simeq}3.17/4/\omega_0^2[E_0\exp(-t^2/T^2)]^2$,
suggesting that the pulse duration can be extracted by fitting the
HCOs with a Gauss squared function. We devise a simple algorithm
to deal with it. Firstly, normalize $\textrm{HCO}-I_p$ by its
maximum, consequently $(\textrm{HCO}-I_p)_N$ obeys the
Gauss-squared function $exp[-(t/T)^2]^2$, which is then
transformed as $\{-ln[(\textrm{HCO}-I_p)_N]/2\}^{0.5}\sim{t/T}$.
The parameter T can be retrieved with the least squares fitting
method. In Fig. \ref{fig6}(a), we shows the
$\{-ln[(\textrm{HCO}-I_p)_N]/2\}^{0.5}$ for a 5- (circles), 8-
(triangles) and 11.5-fs (squares) pulses, which are fitted by the
red, green and blue lines, respectively. The slopes are 0.6016,
0.3688, 0.2666, and the parameter T is retrieved as 1.66, 2.71,
3.75 $T_0$, corresponding to 5.18, 8.5, 11.77 fs, respectively,
which all agree well with the exact values. Note that this method
enables one to retrieve the pulse duration without knowing the
CEP. Figure \ref{fig6}(b) shows the
$\{-ln[(\textrm{HCO}-I_p)_N]/2\}^{0.5}$ and the fitted lines for
the 5-, 8-, 11.5-fs pulses with a CEP of $0.5\pi$. The retrieved
pulse durations are 5.48, 8.6 and 12 fs, respectively. We have
also calculated for other CEPs, the deviations are bellow $8\%$,
showing that this method allows one measured the pulse duration
independently on the CEP.

In summary, we investigate the HHG in an ultrashort linearly
polarized laser pulse by mixing a weak uv pulse. It is shown that
the uv pulse enhances and confines the photonionization in a short
interval, and the high harmonics generated in the individual half
cycle are clearly separated in the time domain by a 1-fs uv pulse.
Further, we show that the HCOs depend sensitively on the CEP of
the driving field, enabling one to {\it in situ} retrieve the
driving laser pulse information with a high accuracy. In addition,
we present an alternative way to measure the pulse duration
independently on the CEP, and the deviation of the retrieved pulse
duration is bellow $8\%$. We also investigate the influence of the
fluctuation on our method. The test simulation shows that neither
a variation of the intensity ($5\times10^{12}-1\times10^{14}$
$\textrm{Wcm}^{-2}$) nor a variation of wavelength (80-150 nm) of
the uv pulse change the above results significantly. Moreover, we
can clearly distinguish the HCH for a range of laser intensities
of $1-6\times10^{14}$ $\textrm{Wcm}^{-2}$ and a broader range is
possible by taking other target atoms (e.g., He$^+$, Ne). The HCOs
still are distinguishable when the pulse duration is increased to
15 fs, indicating a potential to CEP measurement for a multi-cycle
pulse. For even longer pulse, the HCHs still are separated in the
time domain, but the HCO shift becomes more smaller and more
difficult to measure.

\section*{Acknowledgements}
This work was supported by the National Natural Science Foundation
of China under grant Nos. 10574050, 10774054, 10734080 and 973
program under grant No. 2006CB806006. $^\dag$ Author to whom
correspondence should be addressed.

\end{document}